# Comparative Silane Surface Functionalization Strategies for Enhanced Bloch Surface Wave Biosensing of Anti-SARS-CoV-2 Antibodies

**Agostino Occhicone**[a,b,†], **Alberto Sinibaldi**[a,b,†], **Paola Di Matteo**[a], **Daniele Chiappetta**[a,b], **Riccardo Guadagnoli**[a,‡], **Peter Munzert**[c], **and Francesco Michelotti**[a]

[a] *Department of Basic and Applied Science for Engineering, Sapienza University of Rome, Via A. Scarpa 16, 00161 Rome, Italy.*

[b] *Italian Institute of Technology (IIT), Center for Life Nano and Neuro Science, Viale Regina Elena 291, 00161, Rome, Italy.*

[c] *Fraunhofer Institute for Applied Optics and Precision Engineering IOF, Albert-Einstein-Str. 7, Jena 07745, Germany.*

[†] *The authors contributed equally to this work.*

[‡] *Present address: CIC biomaGUNE, Basque Research and Technology Alliance (BRTA), Paseo de Miramón 194, Donostia-San Sebastián 20014, Spain.*

**Corresponding authors:** Alberto Sinibaldi, alberto.sinibaldi@uniroma1.it

**Abstract:** Surface functionalization plays a decisive role in the performance of biosensors, as it governs the efficiency and stability of biomolecule immobilization at the sensor interface and, consequently, the overall performance of the biosensing platforms. In this work, we present a comparative study of three organosilane chemistries — APTES, APDMS, and CPTES — applied to a $SiO_2$-terminated 1D photonic crystal able to sustain Bloch surface waves and designed to operate as optical biosensors in both label-free and fluorescence-enhanced modes. Each chemistry was evaluated through a standardized label-free protocol based on the interaction between immobilized SARS-CoV-2 spike protein and its corresponding antibodies, enabling quantitative assessment of binding efficiency, nonspecific adsorption, and signal repeatability. CPTES exhibited the most favorable balance between specific signals, reduced variability, and low nonspecific adsorption. The three chemistries were subsequently tested in fluorescence mode for the detection of anti-SARS-CoV-2 IgG antibodies in human serum, demonstrating the suitability of BSW-enhanced fluorescence for rapid serological analysis. Overall, the study identifies CPTES as the most robust and reproducible functionalization strategy among the three investigated for BSW biosensing and highlights the potential of the platform for fast, sensitive detection of clinically relevant antibodies.

# 1. Introduction

In the fields of medicine and pharmacology, optical biosensors are widely used devices for the diagnosis of diseases [1]. Label-based biosensors, such as fluorescence biosensors, represent the first choice for highly sensitive detection of biological molecules [2]. In order to overcome the limits of fluorescence biosensors, such as the photobleaching of the fluorophores and the alteration of the biological molecules involved in detection, many advances in the field of label-free biosensors have been made, promoting Surface Plasmon Polariton (SPP) based biosensors as some of the best label-free detection devices [2] [3] [4]. In recent years, biosensors based on the propagation of electromagnetic surface waves on the truncated edge of a one-dimensional photonic crystal (1DPC) have been proposed as a valid alternative to the SPP based biosensors [5]. The main advantage of this detection mechanism is the possibility to tune the properties of the sensor by choosing the materials composing the multilayer and their thicknesses, leading to a better confinement of the electromagnetic waves at the surface [6].

Recently, scientific research has focused on the global health emergency caused by the SARS-CoV-2 virus, achieving rapid and comprehensive characterization of its genomic sequence within a remarkably short timeframe [7]. This unprecedented effort has, in turn, enabled the accelerated development and deployment of sensitive and specific diagnostic tools [8] [9]. To investigate the immune response to SARS-CoV-2, it is essential to assess the presence and activity of neutralizing antibodies. This can be achieved through serological testing, commonly performed using immunoassays, which enable the sensitive and specific detection of antibody-antigen interactions [10] [11].

The successful implementation of an immunoassay requires the stable and functional immobilization of bio-recognition molecules onto the biosensor surface. Surface functionalization is thus a critical step in biosensor development, as it directly influences sensitivity, specificity, and overall assay performance [12]. In particular, the chemical modification of inorganic transducer interfaces to promote efficient, oriented, and stable immobilization of biological recognition elements remains a central challenge and a key area of ongoing research in both academia and industry.

Organosilane-based chemical agents are widely employed in biosensing strategies to modify the surface chemistry of transducers in a broad range of biosensor applications [13], with relevance in optical biosensors [14] [15]. The most widely used functionalization protocol in our group relies on APTES (3-aminopropyltriethoxysilane) [16], previously adopted in several studies [15] [17] [18]. In the present work, alongside APTES, we evaluated two additional organosilanes: APDMS (3-(ethoxydimethylsilyl)propylamine) [19], and a chlorosilane, CPTES (2-chloroethyltriethoxysilane) [20] [21]. These compounds are typically allowed to react with the surface of the biosensors in anhydrous solvents, forming a thin functional layer that enables subsequent bio-functionalization. Moreover, they are known for their high reactivity with various substrates, including silicon oxide ($SiO_2$), which serves as the top layer of the 1DPC biosensor used in the experiments presented in this work [14] [21].

It is well established that the specific conditions of this initial surface modification step can have a critical impact on the efficiency of subsequent bio-conjugation processes and, ultimately, on the reliability and quality of the biosensing experiments. The APTES surface chemistry procedure was optimized by Gunda et al. [22]. APTES is an aminosilane with 3 ethoxy groups bounded to the silicon atom (Figure 1). The APDMS surface chemistry procedure was optimized by Hernandez et al. [19] and it is an aminosilane with two methyl groups and one ethoxy group bounded to the silicon atom (Figure 1). Functionalization with CPTES for the silanization of a $SiO_2$ layer has been explored in some studies such as [23] [24] [25]. During silanization, silane molecules adsorb onto the hydrated $SiO_2$ surface, where the hydrolysis of the silane groups is promoted [26]. This leads to the covalent immobilization of Self-Assembled Monolayers (SAMs) of organosilanes [27].

In the present work, the 1DPC was designed and fabricated to support Bloch surface waves (BSWs) [28] [29] at the interface between the multilayer structure and the biological sample, enabling operation in both label-free (LF) and enhanced fluorescence modes (FLR) [17] [30] (the 1DPC structure is described in the "5.1 Photonic crystal design and fabrication" section). The LF mode exploit a transverse electric (TE)-polarized BSW at the wavelength $\lambda_{LF} = 670\ nm$ and allows real-time monitoring of the biomolecular interaction between a probe molecule and its specific antibody via resonant excitation [31] [32] [33]. In contrast, the FLR-based mode uses a transverse magnetic (TM)-polarized laser diode operating at $\lambda_{FLR} = 405\ nm$ to resonantly excite fluorescent molecules located on the top surface of the 1DPC. The FLR-based mode offers high sensitivity, making it particularly suited for detecting low concentrations of biomarkers in human fluids [15] [18].

In this work, an optical setup which makes use of BSW based biosensors for the detection of Anti-SARS-CoV-2 related antibodies in human serum is proposed. The platform provides results with response rapidity (∼30 minutes) [15], thus positioning itself as a strong competitor to current SARS-CoV-2 serological detection techniques, such as ELISA and CLIA tests, which are accurate but require a laboratory to be performed, and the lateral flow immunoassays (LFIAs), which are rapid but lack sensitivity [9] [34].

The chemistries comparison is methodically carried out using the LF signal shifts induced by molecular interactions with SARS-CoV-2-related proteins. Specifically, the comparison focused on evaluating the interaction strength between the wild type RBD Spyke (Swt) protein bounded at the top surface of the 1DPC and its corresponding Anti-S antibodies [35]. Finally, the three chemistries are used for detecting anti-SARS-CoV-2 IgG antibodies in human serum samples in both LF and FLR mode. The main objective is to fine-tune the biointerface to enable the detection of IgG antibodies in serum samples from COVID-19 patients. This study builds upon our previously published works on the detection of anti-SARS-CoV-2 IgG and IgM antibodies [15] [20].

## 2. Surface bio-activation of the 1DPC-based biochip

### 2.1. Organosilane chemistries

The procedure used for silanization is schematically sketched in Figure 1 and all the details related to the chemical and biological regents are reported in section "5.3 Chemical and biological reagents". The first step (I, surface cleaning step in Figure 1) of the functionalization procedures consists in cleaning of the 1DPC surface using piranha solution (3:1, $H_2SO_4/H_2O_2$). Small amounts of piranha solution are gently deposited for 10 minutes in the area of the 1DPC relating to the two fluidic channels used for two separate experiments. This mixture permits to remove every unwanted molecules and allows the hydroxylation of the surface, making it more hydrophilic. Then, the substrates are rinsed with de-ionized water (DI-$H_2O$) and dried. Drying must be complete, because organosilanes do not polymerize correctly in the presence of water, producing an inhomogeneous layer.

After cleaning, in the following step, the silanization takes place and the SAM is formed (II, SAM step in Figure 1). In the APTES case, the substrates are immersed for 1 hour in a 2% solution of APTES in ethanol. In the APDMS case, the chips are immersed for 1 hour in a 2% solution of APDMS in toluene. In the CPTES case, the substrates are immersed for an hour in a 2% CPTES + 1% N,N-diisopropylethylamine (DIPEA) solution in toluene. The DIPEA nitrogen atom has a lone pair which can react only with small electrophiles (such as protons), because the two isopropyl groups and the ethyl group occupy a lot of space around the N atoms [36]. DIPEA reacts with the OH groups on the surfaces of the chip and subtracts the hydrogen atom from them. In this way, the silicon atoms can covalently bond the oxygen atoms and the bonding between the chlorine atoms and the OH groups are avoided.

Successively (III, curing process step in Figure 1), the substrates are sonicated 3 times for 30 seconds in ethanol in the APTES case, in toluene in the other cases, with a 30 second pause between each sonication. In this passage, residual bubbles between the silane chains are removed, along with the impurities, and the monolayer is relaxed. After that, the chip is rinsed with DI-$H_2O$ and dried. Then, the substrates are heated up between 50°C and 60°C on a hot plate, for 1 hour. This permits to decrease the residual tension of the SAM and to have a vertical preferential orientation, ensuring that the amine group is exposed and strengthening the Si-O-Si bonds.

The final step foresees the activation of the surface (IV, activation step in Figure 1). The amine group of the APTES molecule is activated to improve the coupling of the antibody to the silanized surface by means of crosslinking with glutaraldehyde (GAH). The substrate is immersed in a saline buffer solution 0.1 M NaHCO$_3$, containing 2 mM sodium cyanoborohydride, with 1% of GAH, for 1 hour [22]. GAH's aldehydic group reacts with the amine group exposed by the SAM forming an amide bond, while the other aldehydic group can react with another biological species, such as capture antibodies. Hence, in this final step a silanized surface capable of immobilizing specific molecules is obtained. In the APDMS functionalization protocol, the used cross-linker is the carbonyldiimidazole (CDI). This compound has two imidazole carbamate functional groups and reacts with the amine group of the APDMS using one of them, thus forming a stable carbamate linkage [19]. CDI is easy to use in water solution (150 mM) and successful results have been reported in literature [19]. Instead, CPTES is a chlorosilane and, differently from APTES, has a chlorine atom instead of the amine group. This allows for a direct nucleophilic substitution reaction to take place, i.e., the chlorine atoms of CPTES are the leaving groups and electrophilic centers of reaction, while the other groups present on the biomolecule to be immobilized (such as amines, thiols, etc.) are the nucleophilic groups [24]. This means that, unlike APTES and APDMS, no activation process with additional coupling molecules is required. Finally, the substrates are rinsed with DI-H$_2$O, dried and stored in a vacuum chamber, where they remain at least 12 hours before performing an assay.

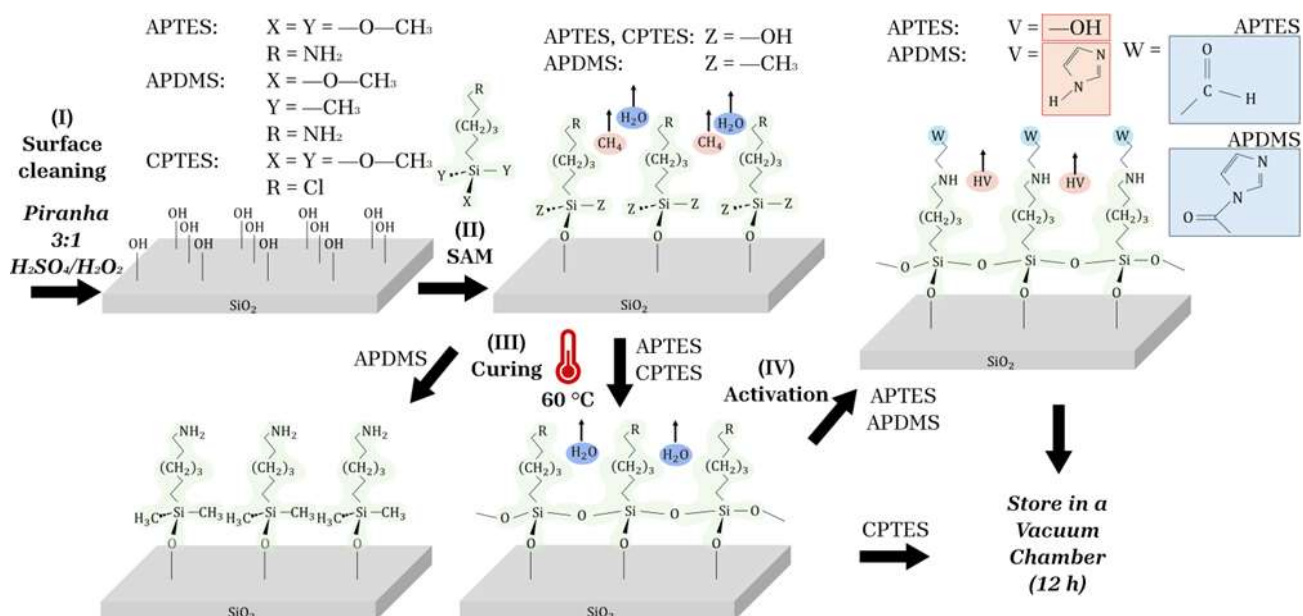

*Figure 1. Schematic representation of the silanization process. The procedure comprises four main steps, which vary depending on the chemistry. (I) Surface cleaning with piranha solution increases the density of surface —OH groups. (II) During SAM formation, the chip is immersed in solutions of APTES, APDMS, or CPTES; simplified molecular structure is shown. (III) In the curing step, the substrate is heated on a hot plate at 50–60 °C for 1 h. (IV) The activation step involves crosslinking with GAH for APTES and with CDI for APDMS; this step is not required for CPTES. After functionalization, substrates are stored in a vacuum chamber for at least 12 h before performing the assay.*

## 2.2. Bio-conjugation of molecular probes

The 1DPC were bio-conjugated with various SARS-CoV-2 related proteins solution (100 μg/mL concentrated) through a nanoplotter (GeSiM). Each biochip contained six or seven (depending by the experimental protocol shown below) discrete sensing regions (Sig), each measuring approximately 1000 × 2000 μm$^2$ and corresponding to a volume of ~600 nL. These regions were plotted along the illumination laser path within each microfluidic channel defined on the 1DPC surface, as shown in Figure 2(a).

The nanoplotter deposition chamber is equipped with shaped aluminum plate that can hold the chips. The aluminum plate is equipped with a water circulation system that enables the temperature to be regulated, and in this instance, it was set at 16°C. Finally, in order to prevent evaporation of the microdroplets deposited on the biochips, the deposition chamber of the nanoplotter is maintained at a relative humidity value of 75% by means of a humidifying system.

After 1 h of incubation time, the chips were taken from the nanoplotter, they were rinsed with DI-H$_2$O and dried, in order to remove the excess liquid. After this passage, they were coupled with a by-adhesive layer (3M™ 468MP Transfer Adhesive Tape) used to define the fluidic cells and mounted on the optical setup (details in "5.2 Optical setup" section and elsewhere in the literature [20]), ready for the measurements. With reference to Figure 12, the acquisition arm is equipped with a cylindrical Fourier imaging system (FL lens for LF mode and FL+ZOOM lenses for FLR mode) and two cylindrical lenses for spotted regions imaging. Light emitted from the biochips is collected and imaged onto a CCD camera. The long axis of the CCD sensor (12 mm, 3388 pixels) is used to resolve the angular components ($\theta$, with $\Delta\theta$ equal to 2.7 and 8.0 deg for LF and FLR mode, respectively), whereas the short axis (10 mm, 2700 pixels) enables the identification of different spatial positions ($\xi$, $\Delta\xi \sim 14\ mm$) along the biochip. As exemplary case, in Figure 2(b-d), the images acquired with the CCD camera at the assay starting time, $t_{exp}^{LF} = 0$, are shown for APTES, APDMS and CPTES surface chemistry, respectively. The seven spotted macro-zones can be clearly distinguished; in fact, the spotted regions show a larger BSW resonance angle. In the map of Figure 2(b), it is superimposed the average reflectance profile measured in the Sig regions.

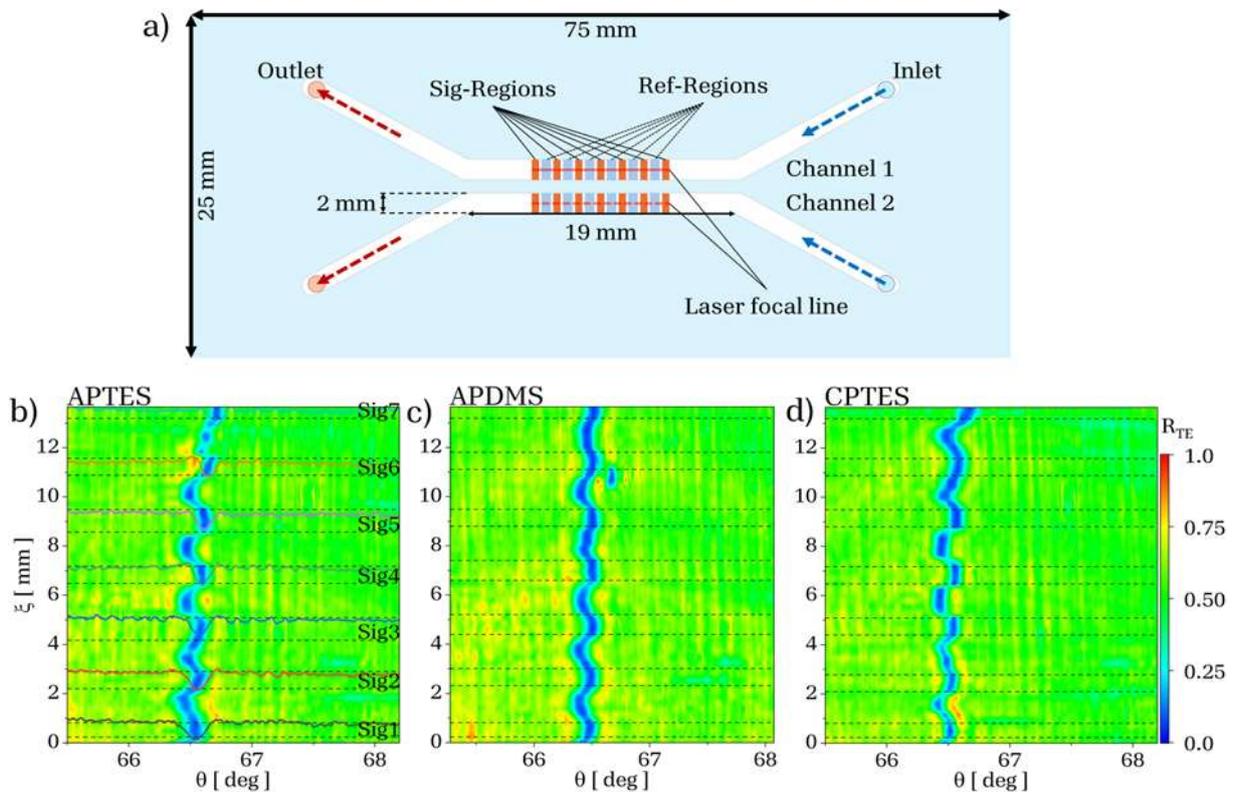

*Figure 2. (a) Sketch of a microscope slide with protein solutions during incubation. (b-d) CCD camera images acquired in LF mode at the beginning of the assays for APTES, APDMS and CPTES, respectively. Overlaid to the images is the spot mask used to identify the different regions on the 1DPC surface from Sig1 to Sig7 (dashed lines). Superimposed to the image (b), we plot the average reflectance profile in the seven different regions. The dip witnesses the excitation of the BSW resonance.*

## 3.    Results and discussion

For each functionalized biochip, the assay is repeated twice by exploiting the two channels used to define the fluidic cells, as shown in Figure 2(a). To compare the different chemistries, a LF assay protocol was implemented (LF-protocol), consisting of a sequence of precisely timed steps, as illustrated in the sensorgrams shown in Figure 3(a-c). Such LF-protocol is limited to LF detection and makes use of synthetic Anti-S antibodies. We also defined a second assay protocol that includes the acquisition of a fluorescence signal. In particular, it includes both LF and FLR measurements, and the use of human sera (HS-protocol).

## 3.1. LF-protocol

Each assay consisted of a series of sequential injections of different solutions in the biochip flow cell while recording the angular position of the BSW resonance. Prior to each assay the tubing and flow cell were primed with the running buffer (PBS 1×). All injections were delivered by a motorized syringe pump (Cavro Centris Pump, Tecan) at a flow rate of $\Phi = 1.37\ \mu L/s$. Each injection was followed by a static incubation of 10 min and a washing step of ~5 min with PBS 1× to remove unbound species.

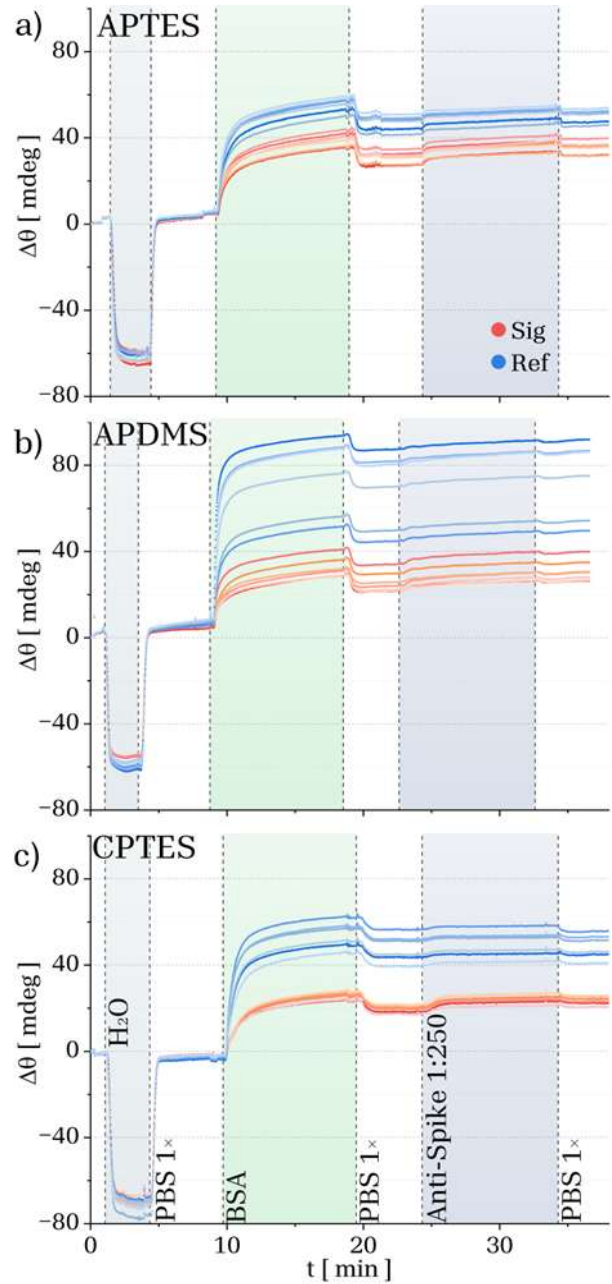

*Figure 3. LF sensograms ($\Delta\theta$ vs time) obtained using the LF-Protocol for the three surface chemistries: (a) APTES, (b) APDMS, and (c) CPTES. The three different sequential injections are highlighted with three different colours: the DI-H$_2$O (grey), the BSA diluted in PBS 1× at 1 mg/mL (green) and Anti-S 1:250 diluted solution (blue).*

At the start of each assay carried out with the LF-protocol, the biochip was filled with DI-H$_2$O and then with PBS 1× to remove non-covalently bound species and to locally evaluate (spot-by-spot) the biochip surface homogeneity. After a 5 min equilibration in PBS 1×, bovine serum albumin (180 $\mu L$ of BSA diluted in PBS 1× at 1 $mg/mL$) was injected at $t_{BSA}$ and incubated for 10 min to block all sites that were not saturated when immobilizing the probes. After washing, synthetic anti-S antibody solution (SARS-CoV-2 Spike monoclonal antibody from SinoBiological, total volume of 105 μL used diluted 1:250 in PBS 1× [37]) was injected at $t_{Ab}$ and incubated for 10 min, followed by a final wash with 180 μL of PBS 1×. The total assay time was approximately 35 min.

The angular shifts (Δθ) of the BSW resonance were recorded continuously as a function of time. The sensograms shown in Figure 3(a-c) correspond to one of two replicate experiments performed on the same biochip; the second replicate is reported in section S1 of Supplementary Information (SI). For each type of surface chemistry, we measured the signals in the Sig regions and in the adjacent Ref regions on the biochip. The experimental curves are the average over 8 statistically independent spots of width 150 μm; the minimum resolvable spot size (60 μm) was set by the diffraction limit of the cylindrical spot imaging optics. In Figure 3, for all three different chemistries, the angular shifts due to the different solution injections are clearly distinguishable and are highlighted with different shaded colours.

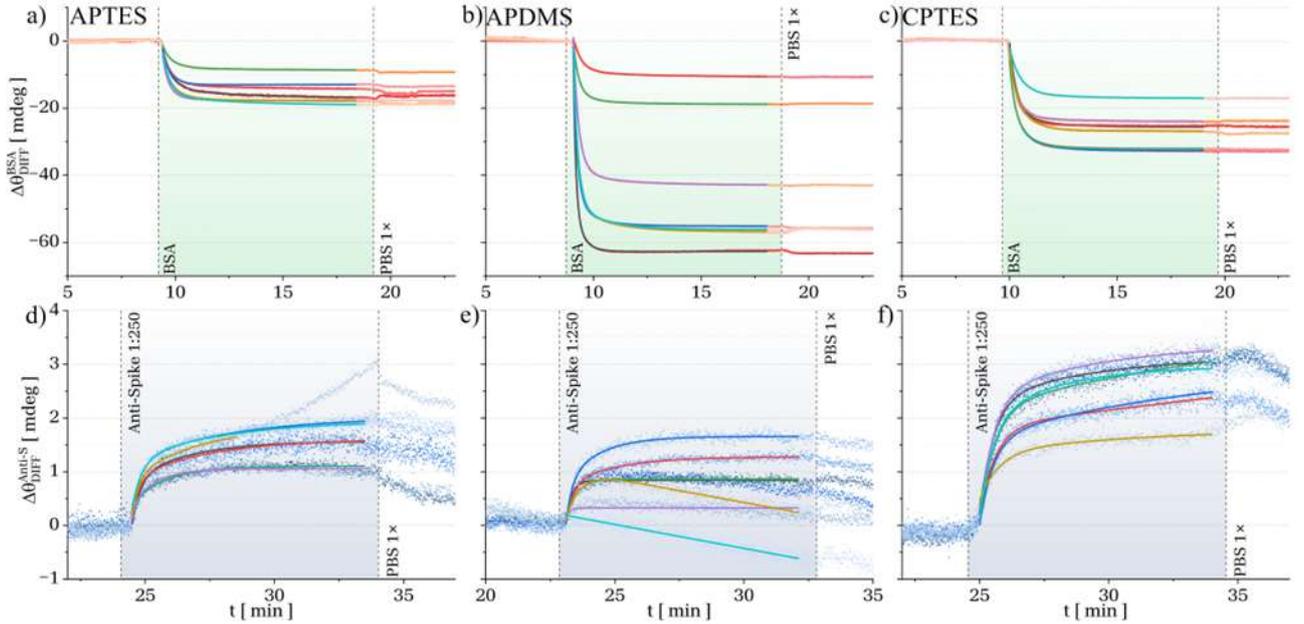

*Figure 4. Panels (a–c) present the differential response to the BSA injection ($\Delta\theta_{DIFF}^{BSA}$) for (a) APTES, (b) APDMS and (c) CPTES. Panels (d–f) show the differential response following the artificial 1:250 Anti-S antibody solution injection ($\Delta\theta_{DIFF}^{Anti-S}$) for (d) APTES, (e) APDMS and (f) CPTES. The solid lines represent the double-exponential fits that best reproduce the experimental data.*

The differential sensograms $\Delta\theta_{DIFF}(t) = \Delta\theta_{Sig}(t) - \Delta\theta_{Ref}(t)$ are shown in Figure 4(a-f). To quantify the net residual angular shift in the Sig regions, the differential sensograms corresponding to the BSA, Figure 4(a–c), and Anti-S, Figure 4(d–f), injection steps were analysed. For each injection, the differential traces were time-shifted so that $\Delta\theta_{DIFF}$ at the injection onset was set to zero and then fitted with a sum of two exponentials (solid lines in the graphs) to extract both the kinetic behaviour and the asymptotic residual angular shift, $\Delta\theta_{DIFF}(t \to \infty) = \Delta\theta_{DIFF}^{\infty}$. BSA injections systematically yielded negative $\Delta\theta_{DIFF}$ values, indicating a higher nonspecific adsorption in the Ref regions compared to the Sig regions. Conversely, Anti-S injections generally produced positive $\Delta\theta_{DIFF}^{\infty}$ values, consistent with specific antibody binding to the spike protein immobilized on the Sig regions. For the APDMS chemistry, some regions exhibited an initial positive response followed by a reduced asymptotic shift, suggesting a limited density of active binding sites.

In the box plot of Figure 5, we summarize the fitted parameters across all Sig regions measured in the two assays performed for each biochip. For each chemistry, we report the $\Delta\theta_{DIFF}^{\infty}$ values (black dots), the mean $\overline{\Delta\theta}_{DIFF}^{\infty}$ (open circle), and the standard error (SE, height of the black box). We observed that APTES, APDMS and CPTES showed negative values of $\overline{\Delta\theta}_{DIFF}^{\infty}$ for BSA injections and positive values of $\overline{\Delta\theta}_{DIFF}^{\infty}$ for Anti-S injections (as shown in Table 1). The thick white whiskers extend to the range defined by the following:

$$\chi_m = Q_1 - 1.5 \cdot IQR$$
$$\chi_M = Q_3 + 1.5 \cdot IQR \qquad (1)$$

where $Q_1$ and $Q_3$ are the first and third quartiles and $(IQR = Q_3 - Q_1)$. Points outside this range are treated as outliers. To combine signal magnitude and repeatability into a single metric we used the following [20]:

$$\chi = \frac{IQR}{\overline{\Delta\theta_{DIFF}^{\infty}}} \qquad (2)$$

Smaller $\chi$ values indicate lower variability relative to the mean. CPTES exhibited the most favourable $\chi$ values for both BSA and Anti-S steps. The two values are highlighted with a yellow background in Table 1, where all the results are also summarized. Finally, in Figure 5, a Kernel Density Estimation (KDE) using a Gaussian kernel was applied to each chemistry's $\Delta\theta_{DIFF}^{\infty}$ distribution to visualize probability density; KDE curves are shown filled in grey, red and blue for APTES, APDMS and CPTES respectively.

*Table 1. Summary of the main results obtained from the residual angular shift $\Delta\theta_{DIFF}^{\infty}$ carried out for the BSA and Anti-S injection steps for the two LF-protocol assays performed for each biochip functionalized with the three chemistries.*

|  | Chem | $n_{SIG}$ | $\overline{\Delta\theta_{DIFF}^{\infty}}$ [mdeg] | SE | Min | $Q_1$ | Median [mdeg] | $Q_3$ | Max | IQR [mdeg] | $\chi$ |
|---|---|---|---|---|---|---|---|---|---|---|---|
| **BSA** | APTES | 14 | -15 | 2 | -36.4 | -17.7 | -13.7 | -11.8 | -2.8 | 5.9 | 0.39 |
|  | APDMS | 14 | -43 | 5 | -62.7 | -56.2 | -52.4 | -19.3 | -10.7 | 36.9 | 0.86 |
|  | CPTES | 14 | -22.9 | 1.8 | -32.7 | -26.8 | -24.6 | -18.1 | -9.9 | 8.7 | **0.38** |
| **Anti-S** | APTES | 14 | 2.6 | 0.4 | 1.1 | 1.6 | 2.0 | 3.7 | 7.1 | 2.1 | 0.81 |
|  | APDMS | 12 | 0.5 | 0.3 | -2.5 | 0.3 | 0.6 | 1.2 | 1.7 | 0.9 | 1.8 |
|  | CPTES | 14 | 3.2 | 0.4 | 0.7 | 2.6 | 3.1 | 3.7 | 6.1 | 1.1 | **0.34** |

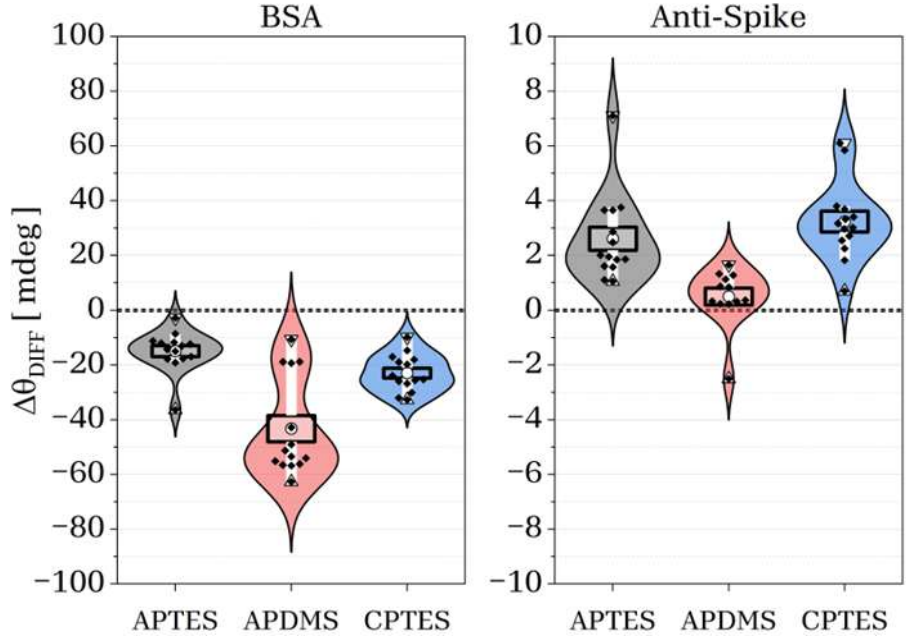

*Figure 5. The boxplot summarizes the main results obtained across all Sig regions measured in the two assays performed for each biochip for both the BSA (left) and Anti-S (right) injections. For each chemistry, we report the $\Delta\theta_{DIFF}^{\infty}$ values (black dots), the mean $\overline{\Delta\theta_{DIFF}^{\infty}}$ (open circle), and the standard error (SE, height of the black box). By applying the interquartile method, the thick white whiskers establish the reliability range of the data. To visualize the probability density, the KDE retrieved using a Gaussian kernel has been plotted for each chemistry's $\Delta\theta_{DIFF}^{\infty}$ measured.*

Moreover, the noise level, $\sigma$, was estimated from the standard deviation of the sensograms during the four minutes immediately preceding the BSA injection (PBS baseline). A confidence threshold of $3\sigma$ was used to discriminate signal from noise (0.37 $mdeg$ for APTES, 0.67 $mdeg$ for APDMS and 0.52 $mdeg$ for CPTES). Because the concentration of Anti-S in solution is unknown, sensitivity, $S$, was expressed in units of angular shift per dilution ratio of Anti-S. Specifically, $S$ was obtained by connecting the origin (0,0) to the measured anti-S point $(4 \cdot 10^{-3}, \overline{\Delta\theta}_{DIFF}^{\infty})$, where $4 \cdot 10^{-3}$ corresponds to the 1:250 dilution. The LF limit of detection (LoD$_{LF}$) in term of dilution ratio units was then estimated as:

$$\text{LoD}_{LF} = \frac{3\sigma}{S} \tag{3}$$

For example, in the case of APTES, the obtained sensitivity is $650 \frac{mdeg}{dilution\ ratio}$ and the LoD can be estimated equal to $0.57 \cdot 10^{-3}$, corresponding to a dilution ratio of about $1:1750$. Using this procedure, we obtained the following LoD$_{LF}$ estimates: $5.4 \cdot 10^{-3}$ (~1:185) for APDMS, and $0.65 \cdot 10^{-3}$ (~1:1540) for CPTES.

### 3.2. HS-protocol

In Figure 6(a), Figure 7(a) and Figure 8(a) we show the results of the assays carried out with the HS-protocol for biochips prepared with the APTES, APDMS and CPTES chemistries, respectively. The assays included fluorescence background (BGD FLR) and signal (FLR) measurements, which were performed before and after the injection of the fluorescent labels, as marked by the vertical bands in the figures. The LF differential sensograms, $\Delta\theta_{DIFF}(t)$, were acquired for the positive (Pos) serum sample; the corresponding assays for the negative (Neg) sample are provided in Section S2 of the SI. For details related to the serum samples, see section "5.4 Biological samples".

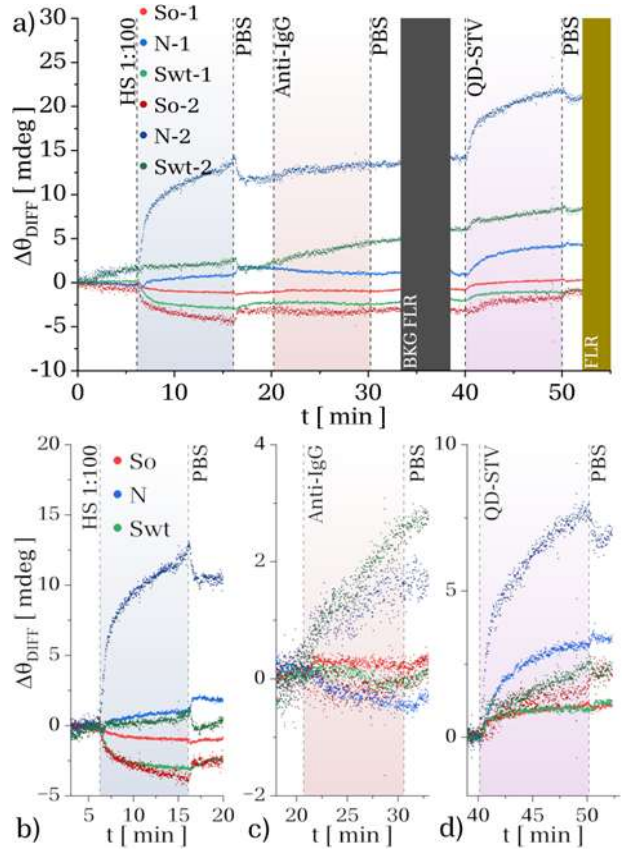

*Figure 6. a) Differential LF sensograms of the six spotted regions of the complete bioassay performed using the HS-protocol on the biochip functionalized with APTES. The injections and incubation phases of the 1:100 diluted HS, Anti-IgG solution, and QD-STV solution are highlighted in blue, red, and violet, respectively. b–d) Enlarged views of the sensogram segments corresponding to the three main injection steps. For clarity, each curve is vertically shifted so that $\Delta\theta_{DIFF} = 0$ at times immediately preceding the respective injections.*

Six signal regions were defined on each biochip, by immobilizing in replicate: Omicron-variant spike proteins (So: Sig1, Sig 4), nucleocapsid proteins (N: Sig2, Sig 5) and wild-type Spike proteins (Swt: Sig3, Sig 6). The figures show the mean LF sensograms, average of 8 spots per region, for So (red), N (blue) and Swt (green). Notably, the So variant had not developed at the time the positive serum samples were collected.

Similarly to the LF-protocol, two initial PBS 1× and DI-H$_2$O washing steps were followed by BSA blocking and rinsing with PBS 1×. The assay begins with the injection of the diluted Pos (or Neg) human serum (105 μL of crude serum diluted 1:100 in PBS 1× containing 0.1% BSA) followed by a 10 min incubation and a subsequent wash with PBS 1×.

Magnified graphs of the human serum injection step ($t = [0, 20]\ min$) are shown in Figure 6(b) for APTES, Figure 7(b) for APDMS and Figure 8(b) for CPTES. The LF differential sensograms do not consistently exhibit homogeneous behavior across homologous regions, likely due to the complexity of the solution matrix; however, a qualitative discrimination between 1:100 diluted positive and negative sera remains at least partially achievable (see Fig. S.2 in SI). However, the 3σ criterion was not consistently satisfied in LF mode for detecting Anti-N and Anti-S antibodies (i.e., IgM and IgG proteins) in the HS samples [38].

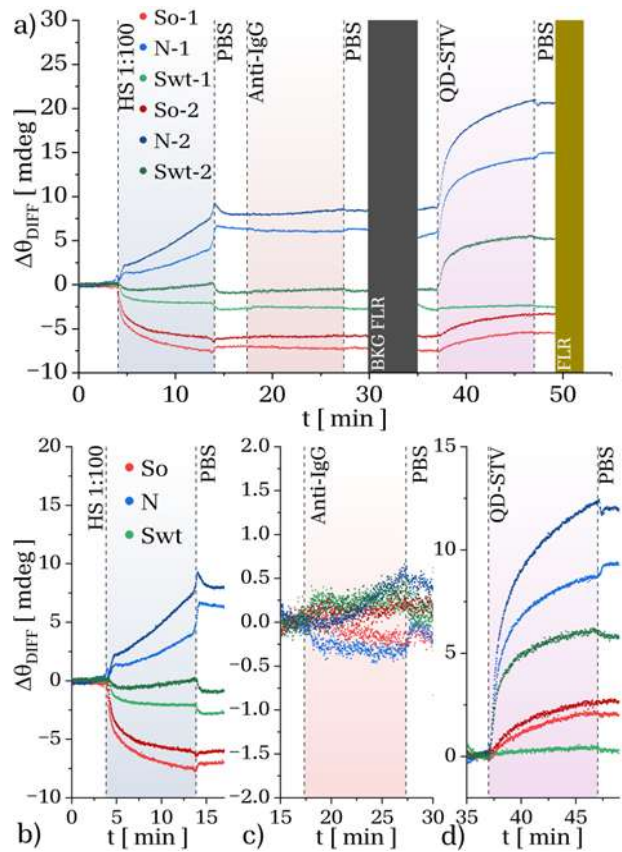

*Figure 7. a) Differential LF sensograms of the six spotted regions of the complete bioassay performed using the HS-protocol on the biochip functionalized with APDMS. The injections and incubation phases of the 1:100 diluted HS, Anti-IgG solution, and QD-STV solution are highlighted in blue, red, and violet, respectively. b–d) Enlarged views of the sensogram segments corresponding to the three main injection steps. For clarity, each curve is vertically shifted so that $\Delta\theta_{DIFF} = 0$ at times immediately preceding the respective injections.*

The FLR part of the HS-protocol includes the following steps. Injection of a solution of biotinylated Anti-IgG detection antibodies (Bio-Plex Pro Human IgG Detection Antibody Anti-SARS-CoV-2 kit from BIORAD), diluted 1:50 in PBS 1× (total volume 105 μL), lower dilution ratio than recommendation reported in the datasheet, followed by incubation for 10 min and a PBS 1× washing step. Injection of 110 μL of streptavidin-conjugated CdSe/ZnS core-shell quantum dots (QD-STV, emission peak at $\lambda_{QD} = 655\ nm$ [39]) at 3 nM in PBS 1×, followed by incubation for 10 min and PBS 1× washing step. The QD-STV emission lies well within the photonic bandgap of the employed 1DPC, whereas the absorption and emission spectra of the QD-STV are reported in the section S3 of the SI, as retrieved using Thermo-Fisher's Spectraviewer [40]. Before QD-STV injection, a fluorescence background (FLR-BKG) is acquired; after incubation, the FLR measurement is performed; both the FLR measurements are performed with the 405 nm laser diode operating at 50 mA. These steps are indicated by the grey and yellow regions in Figure 6(a), Figure 7(a) and Figure 8(a). The magnified graphs of the LF sensograms recorded during such the Anti-IgG and QD-STV injection/incubation steps are shown in Figure 6(c,d) for APTES, Figure 7(c,d) for APDMS and Figure 8(c,d) for CPTES.

The refractive-index perturbation introduced by QD-STV binding at the sensing surface produces a measurable residual angular shift even at the low IgG concentrations present in the diluted Pos serum. IgG is typically the dominant antibody class in the later stages of the immune response, although the relative timing and levels of IgM and IgG production in SARS-CoV-2 infection are known to vary significantly [38] [41]. The large refractive index perturbation produced by the QD captured on the sensitive surface during QD-STV injection enables the platform to operate in an amplified refractometric regime, as previously reported in ref. [15]. The differential sensograms in Figure 6(d), Figure 7(d) and Figure 8(d) clearly show a QD-amplified LF signal. Residual angular shifts following the QD-STV step demonstrate that, even at a 1:100 dilution, Pos and Neg serum samples in the N-protein regions are clearly discriminated for all three chemistries, consistently satisfying the 3σ criterion. By contrast, the response of the spike-protein regions is more heterogeneous and of difficult reading for both Swt and So, preventing a definitive interpretation.

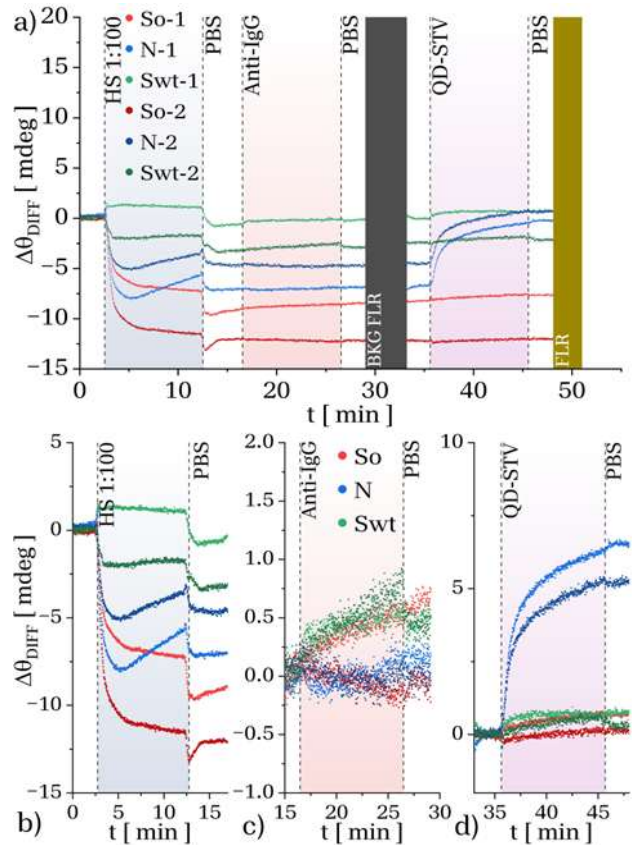

*Figure 8. a) Differential LF sensograms of the six spotted regions of the complete bioassay performed using the HS-protocol on the biochip functionalized with CPTES. The injections and incubation phases of the 1:100 diluted HS, Anti-IgG solution, and QD-STV solution are highlighted in blue, red, and violet, respectively. b–d) Enlarged views of the sensogram segments corresponding to the three main injection steps. For clarity, each curve is vertically shifted so that $\Delta\theta_{DIFF} = 0$ at times immediately preceding the respective injections.*

Figure 9(a,b) present the background subtracted FLR emission maps acquired by the CCD at the conclusion of the assays, following QD incubation and rinsing with running buffer. The maps compare assays performed using Pos, Figure 9(a), and Neg, Figure 9b, human sera. The FLR maps were generated by resonantly exciting the BSW at $\lambda_{exc}$ and detecting the BSW-coupled fluorescence emission at $\lambda_{em}$.

Across all three functionalization chemistries, the FLR signal displays a spatial modulation that correlates precisely with the surface patterning, thereby validating the biochip's selectivity for anti-SARS-CoV-2 antibodies. Under fixed resonant excitation and constant exposure time, the Neg samples consistently exhibited significantly lower signal modulation. This confirms that the biochip can reliably discriminate between Pos and Neg samples (diluted 1:100) via IgG antibody detection, regardless of the specific chemistry employed.

The average FLR emission intensity across the regions functionalized with So, N, and Swt proteins is illustrated in Figure 9(c). The average FLR emissions were processed by fitting with a Lorentzian function and subtracting the retrieved offset values. In all instances involving positive human serum, the mean intensity over the N-functionalized regions (blue dots) is clearly distinguishable from the negative controls (semi-transparent blue dots).

The intensity profiles along the N-functionalized regions confirm the efficient coupling of QD emission to the TE-polarized BSW modes supported by the 1DPC. The fitting parameters for these peaks (center position $x_c$, width, $w$, and height, $I_m$) are summarized in Table 2. For the N-protein, the average intensity difference between Pos and Neh samples ($\overline{\Delta I}_{TE}^{FLR} = \bar{I}_m^{Pos} - \bar{I}_m^{Neg}$) was quantified as 8500, 13900, and 20800 counts for APTES, APDMS, and CPTES, respectively.

A similar evaluation was conducted for the S proteins. For the So variant, the $\overline{\Delta I}_{TE}^{FLR}$ values were 1400 (APTES), -2300 (APDMS), and 4800 (CPTES). For Swt detection, the results were 1200 (APTES), -1400 (APDMS), and 4800 (CPTES). It is important to note that the negative values observed with the APDMS chemistry for S proteins deviate from expected results, despite the samples being collected prior to the emergence of the Omicron variant.

*Table 2. Summary of the main results obtained from the emission spectra collected in the FLR operation mode for the three different chemistries and for each Sig region.*

|  |  | APTES | | APDMS | | CPTES | |
|---|---|---|---|---|---|---|---|
|  |  | POS | NEG | POS | NEG | POS | NEG |
| **So-1** | $x_c$ [deg] | 4.498 ± 0.003 | 4.963 ± 0.005 | 4.609 ± 0.003 | 4.981 ± 0.007 | 5.227 ± 0.003 | 4.634 ± 0.006 |
|  | $w$ [deg] | 0.92 ± 0.01 | 0.99 ± 0.02 | 0.94 ± 0.01 | 0.72 ± 0.03 | 0.96 ± 0.01 | 0.98 ± 0.02 |
|  | $I_m$ [a.u.] | 4357 ± 30 | 3862 ± 39 | 5637 ± 39 | 9850 ± 200 | 5662 ± 34 | 1401 ± 18 |
| **N-1** | $x_c$ [deg] | 4.612 ± 0.003 | 5.012 ± 0.005 | 4.606 ± 0.002 | 4.96 ± 0.01 | 5.224 ± 0.003 | 4.638 ± 0.004 |
|  | $w$ [deg] | 0.953 ± 0.009 | 0.98 ± 0.02 | 0.970 ± 0.009 | 0.89 ± 0.05 | 0.974 ± 0.009 | 0.942 ± 0.013 |
|  | $I_m$ [a.u.] | 13184 ± 72 | 4629 ± 44 | 26580 ± 140 | 11600 ± 300 | 21310 ± 120 | 2487 ± 19 |
| **Swt-1** | $x_c$ [deg] | 4.562 ± 0.003 | 5.007 ± 0.006 | 4.612 ± 0.003 | 4.969 ± 0.013 | 5.253 ± 0.003 | 4.606 ± 0.005 |
|  | $w$ [deg] | 0.92 ± 0.01 | 0.95 ± 0.02 | 0.896 ± 0.009 | 0.90 ± 0.06 | 0.95 ± 0.01 | 1.017 ± 0.016 |
|  | $I_m$ [a.u.] | 6067 ± 41 | 4125 ± 49 | 5754 ± 34 | 9600 ± 300 | 6840 ± 46 | 1766 ± 16 |
| **So-2** | $x_c$ [deg] | 4.549 ± 0.003 | 5.018 ± 0.004 | 4.574 ± 0.002 | 5.01 ± 0.01 | 5.190 ± 0.003 | 4.615 ± 0.003 |
|  | $w$ [deg] | 0.95 ± 0.01 | 0.984 ± 0.014 | 0.936 ± 0.008 | 0.92 ± 0.05 | 0.955 ± 0.009 | 0.99 ± 0.01 |
|  | $I_m$ [a.u.] | 9364 ± 61 | 7063 ± 58 | 9621 ± 46 | 10100 ± 300 | 8810 ± 50 | 3545 ± 22 |
| **N-2** | $x_c$ [deg] | 4.663 ± 0.003 | 5.009 ± 0.005 | 4.609 ± 0.002 | 5.01 ± 0.01 | 5.218 ± 0.003 | 4.636 ± 0.003 |
|  | $w$ [deg] | 0.95 ± 0.01 | 1.026 ± 0.02 | 0.979 ± 0.008 | 0.92 ± 0.05 | 1.000 ± 0.009 | 0.98 ± 0.01 |
|  | $I_m$ [a.u.] | 14682 ± 86 | 6264 ± 57 | 23022 ± 110 | 10200 ± 260 | 26230 ± 140 | 3471 ± 21 |
| **Swt-2** | $x_c$ [deg] | 4.634 ± 0.005 | 4.899 ± 0.006 | 4.620 ± 0.003 | 5.03 ± 0.01 | 5.180 ± 0.004 | 4.640 ± 0.007 |
|  | $w$ [deg] | 0.901 ± 0.016 | 0.95 ± 0.02 | 0.957 ± 0.009 | 0.82 ± 0.05 | 1.012 ± 0.013 | 0.89 ± 0.02 |
|  | $I_m$ [a.u.] | 4224 ± 43 | 3839 ± 46 | 7323 ± 41 | 6210 ± 175 | 5717 ± 43 | 1208 ± 20 |

The main results of this comparative analysis are summarized in the column plot of Figure 10, which reports the FLR intensity peaks together with absolute error bars for each surface chemistry and sample type; the width of each column is proportional to the relative error measured in the corresponding spotted region.

A clear correlation emerges between the FLR performance, and the behaviour observed in the differential LF sensograms. As LF measurements probe the effective refractive-index perturbation induced by bio-molecular binding at the sensor surface, weak, heterogeneous or even negative LF responses point to a limited density of active binding sites and/or an increased contribution of nonspecific interactions at the bio-interface. These effects directly impact the subsequent steps of the FLR assay, by reducing the number of captured detection antibodies and, consequently, the amount of QD-STV efficiently immobilized within the BSW evanescent field. Accordingly, surface chemistries characterized by less stable or poorly reproducible LF responses exhibit a reduced FLR contrast and a larger variability between Pos and Neg samples. Overall, CPTES — showing the most robust and reproducible LF behaviour — enables a more efficient build-up of the fluorescent layer and yields the highest signal-to-noise ratio and the most consistent Pos/Neg discrimination in FLR mode. Although a direct quantitative correspondence between LF and FLR responses cannot be established, the data support the view that differences in bio-interface quality, as probed in LF mode, are reflected in the overall efficiency of the FLR readout.

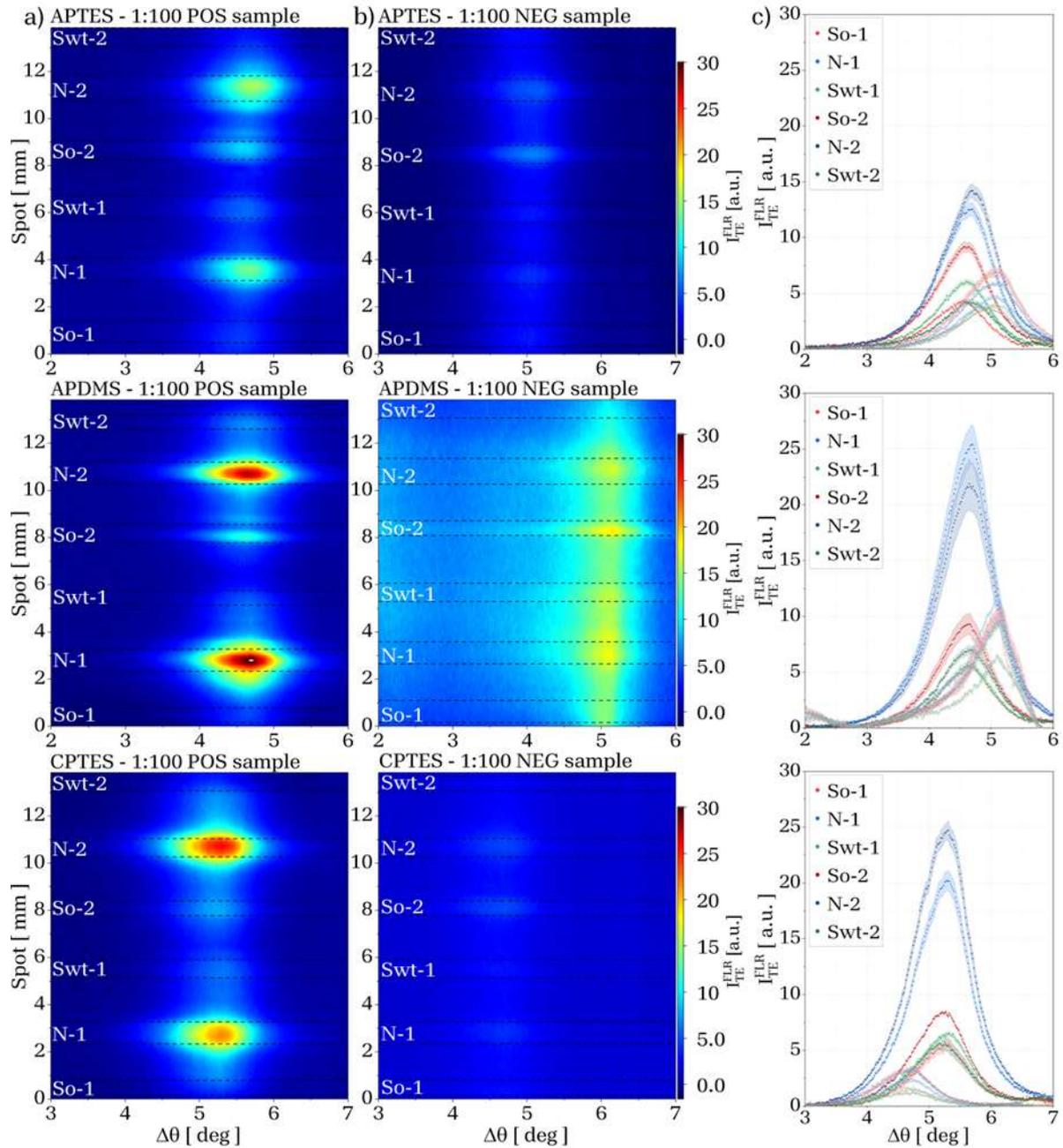

*Figure 9. FLR maps recorded under the same conditions at the end of assays carried out with either positive (a) or negative (b) sera. (c) Average FLR emission intensity from spots in the SIG regions functionalized with So, N, Swt proteins in the case of Pos (darker curves) and Neg (lighter curves) samples.*

## 4. Conclusions

In this work, we systematically compared three organosilane-based surface functionalization strategies — APTES, APDMS, and CPTES — for the bio-interface engineering of a BSW biosensor designed for the detection of SARS-CoV-2–related antibodies. The LF-Protocol enabled a quantitative assessment of nonspecific adsorption, binding efficiency, and signal reproducibility across the three chemistries. Among them, APTES and CPTES demonstrated the most balanced performance, combining low variability, reduced nonspecific binding, and a stable specific response toward Anti-S antibodies with respect to APDMS. These findings were further supported by the sensitivity and LoD estimates, which confirmed the suitability of APTES and CPTES for reliable biosensing applications.

The evaluation of the three chemistries in FLR mode for the detection of anti-SARS-CoV-2 IgG antibodies in human serum highlighted the versatility of the BSW platform, which benefits from enhanced excitation and emission of fluorophores. The combined LF and FLR results demonstrate that the choice of surface chemistry is a critical factor in optimizing biosensor performance. Overall, CPTES emerges as a robust and reproducible functionalization strategy, strengthening the potential of BSW-based biosensors as rapid and sensitive tools for serological diagnostics.

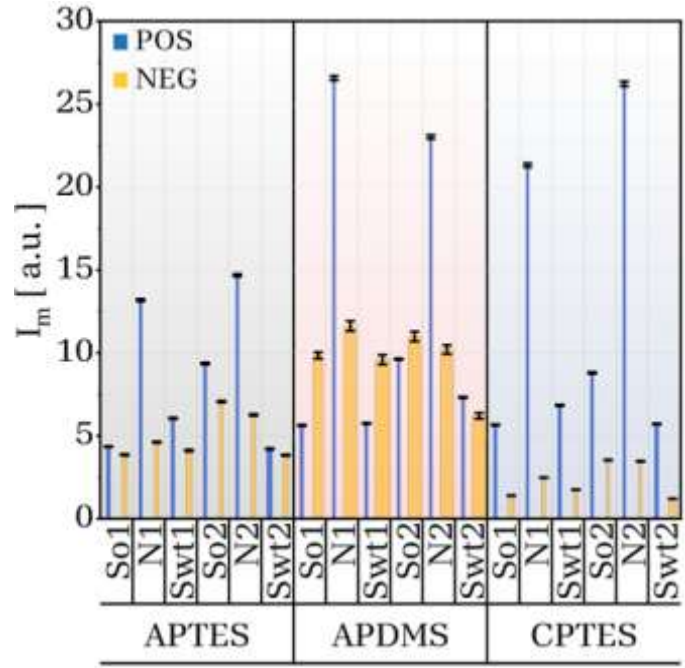

*Figure 10. Maximum FLR intensity column plot for positive (blue columns) and negative (yellow columns) samples for the three different chemistries and for each Sig region. The column thicknesses are proportional to the FLR intensity relative error.*

## 5. Materials and Methods

### 5.1. Photonic crystal design and fabrication

The 1DPC biochips are made up of two alternating layers of silica ($SiO_2$) and tantala ($Ta_2O_5$), the first with thickness $d_{SiO_2} = 275\ nm$ and the second with thickness $d_{Ta_2O_5} = 120\ nm$ [31] [17]. The crystal is covered by a layer of titania ($d_{TiO_2} = 20\ nm$) and a layer of silica ($d_{SiO_2} = 20 nm$), which is necessary in order to work with silane-based functionalization protocols [17]. For $\lambda_{LF} = 670\ nm$, the complex refractive indices are $n_{SiO_2} = 1.474 + i5 \times 10^{-6}$, $n_{Ta_2O_5} = 2.160 + i5 \times 10^{-5}$ and $n_{TiO_2} = 2.28 + i1.8 \times 10^{-3}$. The refractive index distribution in the 1DPC is reported in Figure 11(b).

The 1DPCs were deposited on standard microscope slides, which exhibit a refractive index of $n_P = 1.51$ at $\lambda_{LF}$. The deposition of the dielectric materials was carried out by plasma ion assisted evaporation under high vacuum conditions, using a Syrus Pro 1100 deposition system from Bühler Alzenau GmbH [42].

The 1DPC was designed to operate with external media of refractive index $n_{EM} \approx 1.33$, typical of aqueous solutions [20]. As shown in the reflectance maps of Figure 11(a), the structure supports surface modes in both TE (upper panel) and TM (lower panel) polarizations. We highlight the reflectance profiles at the wavelengths relevant to our experiments: $\lambda_{LF}$, used for LF measurements; $\lambda_{FLR} = 405\ nm$, used for FLR excitation; and $\lambda_{QD} = 655\ nm$, corresponding to the emission of the selected fluorescent marker. Figure 11(b) displays the squared electric $|E|^2$ and magnetic $|H|^2$ field distributions associated with TE and TM BSWs, respectively, when the 1DPC is covered by water. The evanescent field penetrates only a few hundred nanometers into the external medium; using the Transfer Matrix Method [28], we obtained penetration depths $L_{pen}$ of 125 nm and 53 nm at $\lambda_{LF}$ and $\lambda_{FLR}$, respectively.

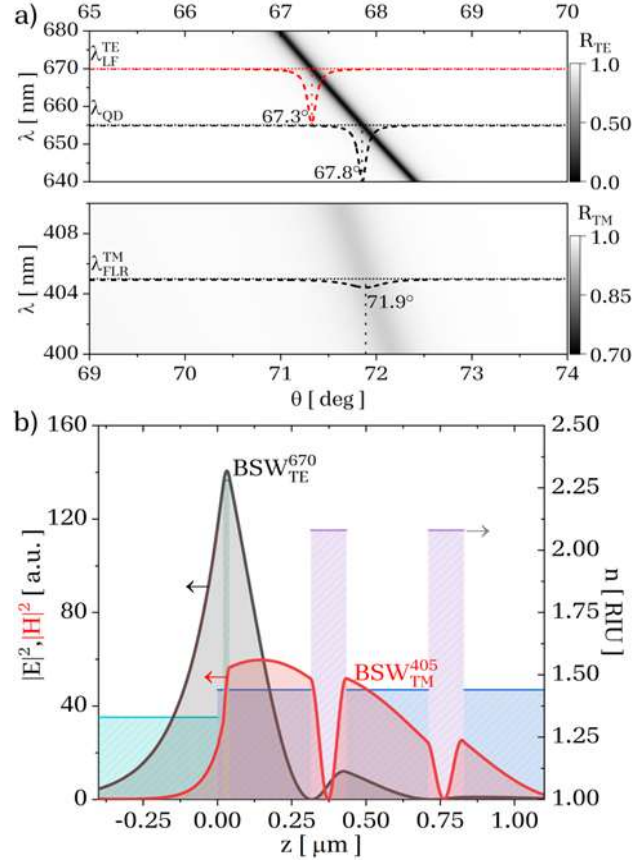

*Figure 11. (a) Reflectance maps for both TE (upper) and TM (lower) polarizations. Superimposed, are shown the reflectance profiles at $\lambda_{LF}$ (red) and $\lambda_{QD}$ (black) in the TE case, and at $\lambda_{FLR}$ in the TM case. (b) 1DPC refractive index distribution with reference to the right axis. Square modulus of electric and magnetic fields (left axis) when BSWs are excited in both TE (black) and TM (red) polarization state at the wavelength $\lambda_{LF}$ and $\lambda_{FLR}$, respectively.*

### 5.2. Optical setup

The optical readout setup is illustrated in Figure 12. The system operates in two distinct configurations and is equipped with two independent CW diode lasers. The first laser diode is used for LF sensing, exciting the BSW at $\lambda_{LF}$ in TE polarization. The second laser is dedicated to FLR excitation by coupling a TM-polarized BSW at $\lambda_{FLR}$, with the resulting enhanced fluorescence collected angularly in TE polarization. The single detection arm, shared by both LF and FLR modes, is equipped with a CCD array (Apogee Ascent, Sony ICX814, 12×10 mm, 3388×2712 pixels). BSW excitation is achieved using a BK7 coupling prism (P) in the Kretschmann–Raether configuration [43]. The 1DPC slide is covered by a microfluidic cell (2 mm wide, 0.2 mm high, 19 mm long), and solutions are delivered via a syringe pump (4-port Cavro® Centris Pump).

The optical components include TE and TM polarizers, cylindrical lenses (CL (f = 150 mm), CL1, CL2, S1, S2), spherical lenses (MO, 40× objective, NA=0.65, f = 3 mm, and SL, f = 250 mm), a dichroic beam splitter (DBS, Chroma ZT640rdc [44]), a half-wave plate (HWP), excitation and emission filters (EXF: Chroma AT405/30x [45]; EMF: Chroma AT450lp [46]), a scattering disk (SD), a Fourier lens (FL, f = 70 mm), and, only in the FLR mode, a cylindrical zoom lens (ZOOM, f = 75 mm).

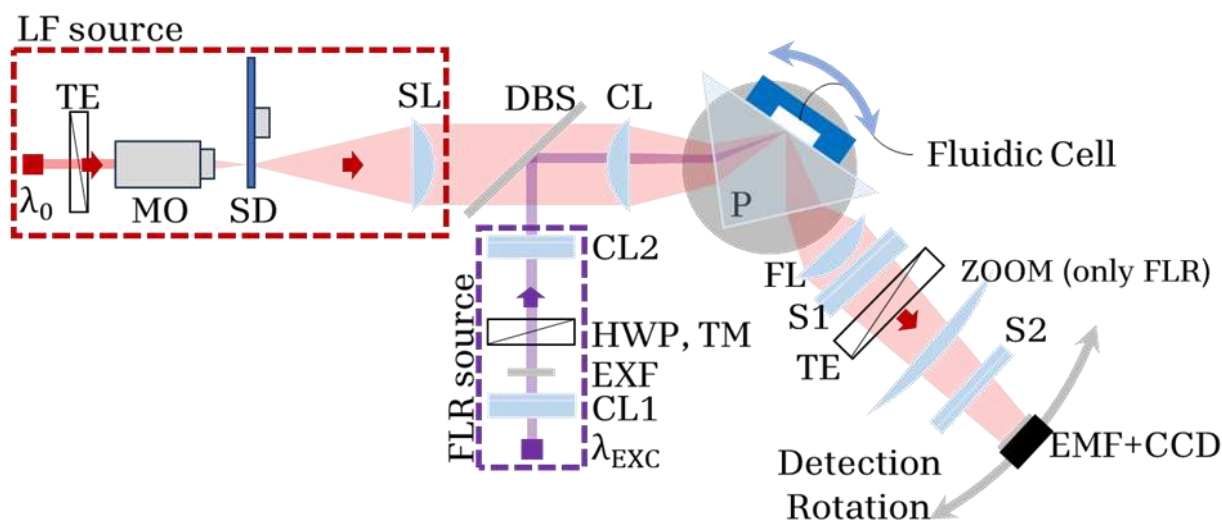

*Figure 12. Sketch of the optical read-out system. TE and TM polarizing filters (TE and TM respectively); cylindrical lenses (CL, CL1, CL2, S1, S2); spherical lenses (MO, SL); dichroic beam splitter (DBS); a half-wave plate (HWP); excitation and emission filters (EXF and EMF, respectively); a scattering disk (SD); a Fourier lens (FL); zoom lens (ZOOM); BK7 prism (P).*

## 5.3. Chemical and biological reagents

APTES, APDMS and CPTES, purity ≥ 97%, were purchased from Sigma Aldrich Merck. GAH stands for glutaraldehyde solution 50 wt. % in $H_2O$ and CDI is a 1,1′-Carbonyldiimidazole in powder with a purity 97% both from Sigma Aldrich Meck. The N,N-diisopropylethylamine (DIPEA), purity 99% was from Thermo Fisher. Ethanol pure ≥99.5%, ACS reagent from Merck. Toluene, purity ≥ 95% was from Carlo Erba. Bovine serum albumin (BSA), fraction V, purity ≥ 98.5%, and the Phosphate Buffered Saline (PBS, pH 7.2-7.6, 10mM) were from Sigma Aldrich Merck. The reagents SARS-CoV-2 Nucleocapsid protein (N), SARS-CoV-2 B1.1.529 Omicron Spike RBD protein (So) and wild-type Spike RBD protein (Swt) were from Sino Biological. Streptavidinated QD655 (QD-STV) were from Thermo Fisher Scientific.

## 5.4. Biological samples

Blood samples were collected from a cohort of representative healthy donors and cancer patients with a prior SARS-CoV-2 infection, sourced from the IRCCS Regina Elena National Cancer Institute (IRE) Biobank. All participants provided written informed consent. Whole blood samples were processed within one hour of collection. Serum was separated by centrifugation at 2000×g for 20 minutes and stored at –80°C in single-use 0.5 mL aliquots, each labeled with a unique, randomly generated barcode.

The concentration of antibodies in each undiluted serum sample was measured using two different chemiluminescence immunoassay (CLIA) systems: the Maglumi 2019-nCoV IgG assay (Snibe; threshold: 1 AU/mL) and the Liaison SARS-CoV-2 S1/S2 IgG assay (Diasorin; threshold: 15 AU/mL). Two serum samples were selected. The Neg sample, with a Maglumi result of 0.1 AU/mL and unavailable Liaison data, was obtained from a donor with no history of SARS-CoV-2 infection. The Pos sample, showing 22.2 AU/mL on Maglumi and 383 AU/mL on Liaison, was collected from a donor who had recovered from the Wuhan original variant of SARS-CoV-2 and tested PCR-negative for at least 86 days prior to sampling. In the assays performed with 1DPC biochips, both Neg and Pos serum samples were tested at a 1:100 dilution in 0.1% BSA prepared in PBS 1×.

## Acknowledgements

This work was supported by the Italian Ministry of Research, under the complementary actions to the PNRR "D34Health - Digital Driven Diagnostics, prognostics and therapeutics for sustainable Health care" Grant # PNC0000001, project code B53C22006120001. The authors acknowledge the contribution of Dr. Matteo

# Comparative Silane Surface Functionalization Strategies for Enhanced Bloch Surface Wave Biosensing of Anti-SARS-CoV-2 Antibodies

**Agostino Occhicone**[a,b,†], **Alberto Sinibaldi**[a,b,†], **Paola Di Matteo**[a], **Daniele Chiappetta**[a,b], **Riccardo Guadagnoli**[a,‡], **Peter Munzert**[c], **and Francesco Michelotti**[a]

[a] Department of Basic and Applied Science for Engineering, Sapienza University of Rome, Via A. Scarpa 16, 00161 Rome, Italy.

[b] Italian Institute of Technology (IIT), Center for Life Nano and Neuro Science, Viale Regina Elena 291, 00161, Rome, Italy.

[c] Fraunhofer Institute for Applied Optics and Precision Engineering IOF, Albert-Einstein-Str. 7, Jena 07745, Germany.

[†] The authors contributed equally to this work.

[‡] Present address: CIC biomaGUNE, Basque Research and Technology Alliance (BRTA), Paseo de Miramón 194, Donostia-San Sebastián 20014, Spain.

**Corresponding author:** Alberto Sinibaldi, alberto.sinibaldi@uniroma1.it

**Abstract:** Surface functionalization plays a decisive role in the performance of biosensors, as it governs the efficiency and stability of biomolecule immobilization at the sensor interface and, consequently, the overall performance of the biosensing platforms. In this work, we present a comparative study of three organosilane chemistries — APTES, APDMS, and CPTES — applied to a $SiO_2$-terminated 1D photonic crystal able to sustain Bloch surface waves and designed to operate as optical biosensors in both label-free and fluorescence-enhanced modes. Each chemistry was evaluated through a standardized label-free protocol based on the interaction between immobilized SARS-CoV-2 spike protein and its corresponding antibodies, enabling quantitative assessment of binding efficiency, nonspecific adsorption, and signal repeatability. CPTES exhibited the most favorable balance between specific signals, reduced variability, and low nonspecific adsorption. The three chemistries were subsequently tested in fluorescence mode for the detection of anti-SARS-CoV-2 IgG antibodies in human serum, demonstrating the suitability of BSW-enhanced fluorescence for rapid serological analysis. Overall, the study identifies CPTES as the most robust and reproducible functionalization strategy among the three investigated for BSW biosensing and highlights the potential of the platform for fast, sensitive detection of clinically relevant antibodies.

**Keywords:** Biosensor platform, Bloch surface waves, surface functionalization, label-free optical sensor, enhanced fluorescence, biomarker.

## S1. LF-Protocol: replicate assay

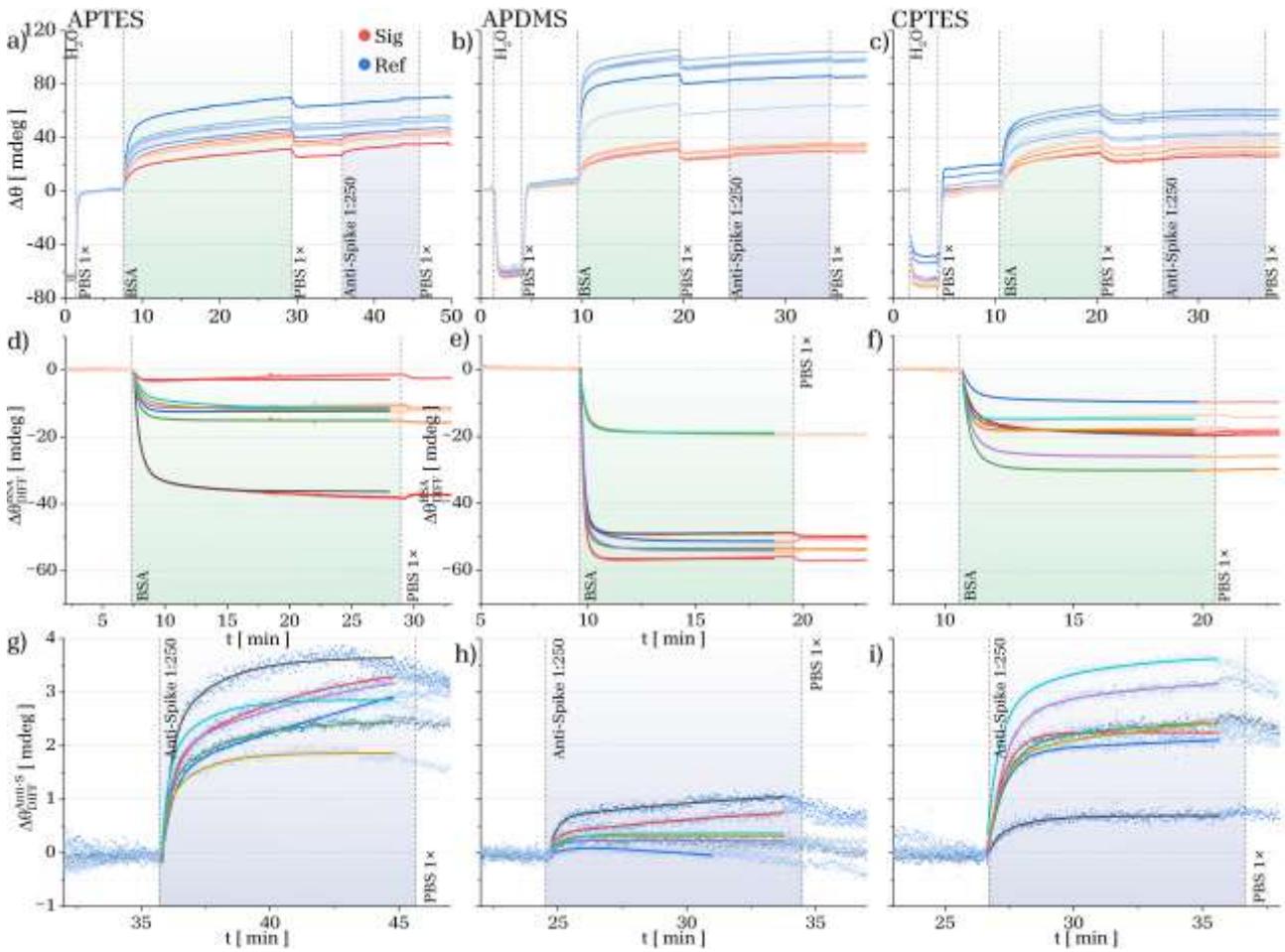

**Figure S.1.** LF ($\Delta\theta$, a-c) sensograms obtained using the LF-Protocol for the three surface chemistries in the second channel of the same biochips shown in the main text: (a) APTES, (b) APDMS, and (c) CPTES. Panels (d–f) present the differential response to the BSA injection ($\Delta\theta_{DIFF}^{BSA}$) for (d) APTES, (e) APDMS and (f) CPTES. Panels (g–i) show the differential response following the artificial 1:250 Anti-S antibody solution injection ($\Delta\theta_{DIFF}^{Anti-S}$) for (g) APTES, (h) APDMS and (i) CPTES. The solid lines represent the double-exponential fits that best reproduce the experimental data.

## S2. HS-Protocol: negative sample

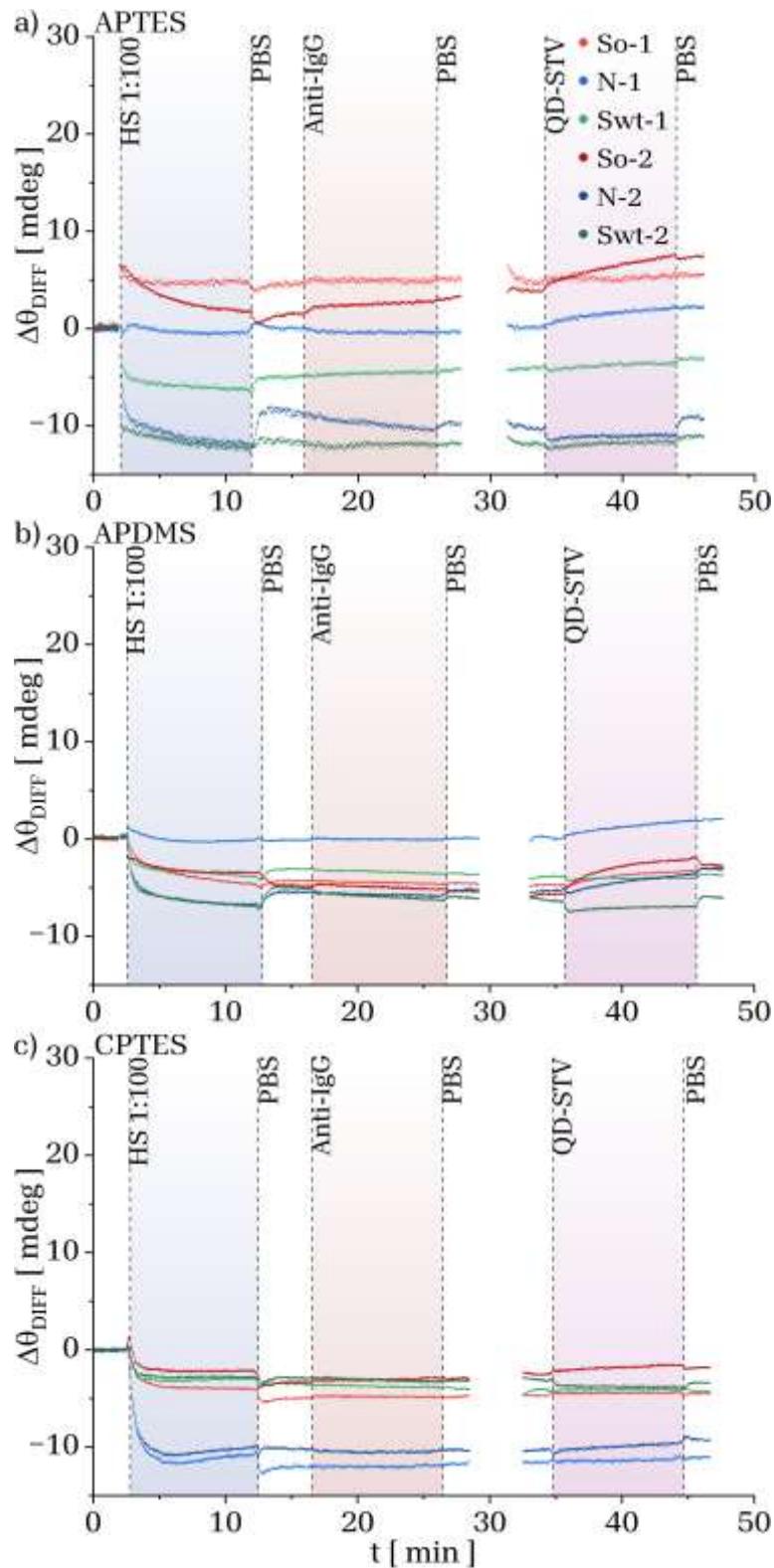

**Figure S.2.** Differential LF sensograms of the six spotted regions of the complete bioassay performed using the HS-protocol on the biochip functionalized with APTES (a), APDMS (b) and CPTES (c). The injections and incubation phases of the 1:100 diluted HS, Anti-IgG solution, and QD-STV solution are highlighted in blue, red, and violet, respectively. For clarity, each curve is vertically shifted so that $\Delta\theta_{DIFF} = 0$ at times immediately preceding the 1:100 diluted HS injections.

**S3.        Absorption and emission spectra of the QD-STV**

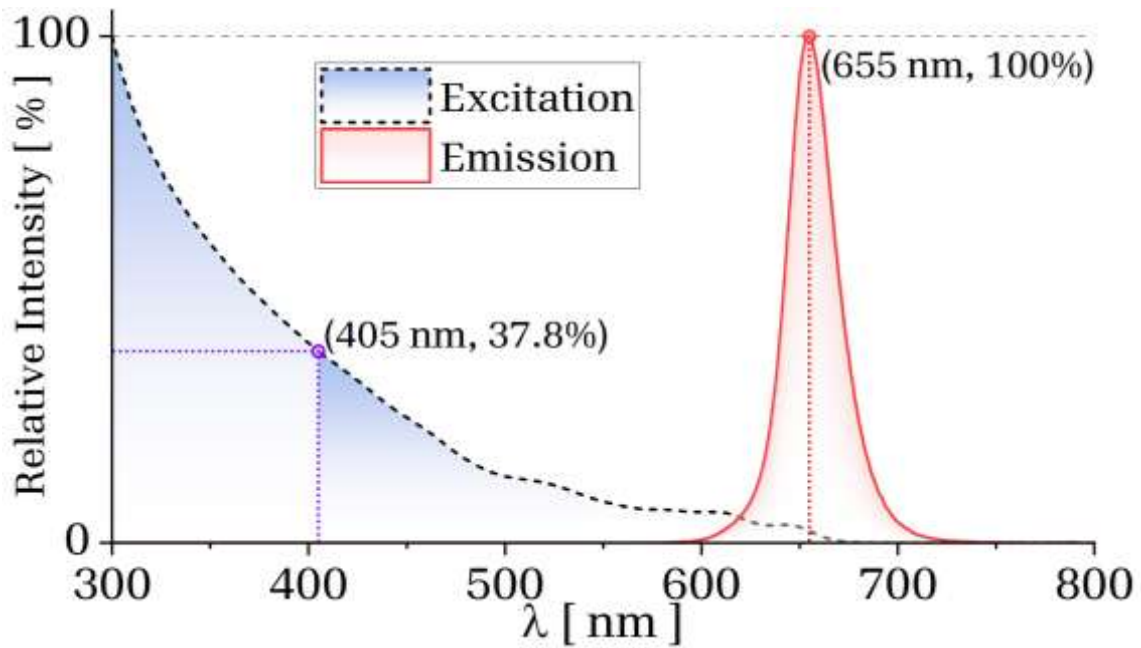

**Figure S.3.** Absorption (black dashed line) and emission (red solid line) spectra of the streptavidin-conjugated CdSe/ZnS core-shell quantum dots (QD-STV) used in the experiments. The QD-STV show an emission peak at $\lambda_{QD} = 655\ nm$ [1]. The data are retrieved from Thermo-Fisher's Spectraviewer [2].